\documentclass[conference, compsoc]{IEEEtran}
\IEEEoverridecommandlockouts
\usepackage{cite}
\usepackage{amsmath,amssymb,amsfonts}
\usepackage{algorithmic}
\usepackage{graphicx}
\usepackage{textcomp}
\usepackage{xcolor}
\usepackage{color}
\usepackage{url}
\usepackage{listings}
\usepackage{color}
\usepackage{hyperref}

\definecolor{dkgreen}{rgb}{0,0.6,0}
\definecolor{gray}{rgb}{0.5,0.5,0.5}
\definecolor{mauve}{rgb}{0.58,0,0.82}
\def\BibTeX{{\rm B\kern-.05em{\sc i\kern-.025em b}\kern-.08em
    T\kern-.1667em\lower.7ex\hbox{E}\kern-.125emX}}

\begin{document}

\title{MVAM: Multi-variant Attacks on Memory for IoT Trust Computing}

\author{
\IEEEauthorblockN{Arup Kumar Sarker\IEEEauthorrefmark{1}, Md Khairul Islam\IEEEauthorrefmark{2}, Yuan Tian\IEEEauthorrefmark{3}}
\IEEEauthorblockA{
\IEEEauthorrefmark{1}\IEEEauthorrefmark{2}University of Virginia, Charlottesville, VA 22904, USA,
\{djy8hg, mi3se\}@virginia.edu}
\IEEEauthorblockA{
\IEEEauthorrefmark{3}University of California, Los Angeles, CA 90095-1405, USA
yuant@ucla.edu}
%\and
%\IEEEauthorblockN{Md Khairul Islam\textsuperscript{2}}
%\IEEEauthorblockA{\textit{mi3se@virginia.edu}}
%\and
%\IEEEauthorblockN{Spencer Thomas Keefer\textsuperscript{3}}
%\IEEEauthorblockA{\textit{kcf7fj@virginia.edu}}
%\and
%\IEEEauthorblockN{Xavier Castillo-Vieira\textsuperscript{3}}
%\IEEEauthorblockA{\textit{xac3cy@virginia.edu}}
%\and

%\IEEEauthorblockN{Yuan Tian\textsuperscript{3}}
%\IEEEauthorblockA{\textit{yuant@ucla.edu}}
% \and
% \IEEEauthorblockN{Felix Xiaozhu Lin\textsuperscript{4}}
% \IEEEauthorblockA{\textit{xl6yq@virginia.edu}}
}

\maketitle

\begin{abstract}

With the significant development of the Internet of Things and low-cost cloud services, the sensory and data processing requirements of IoT systems are continually going up. TrustZone is a hardware-protected Trusted Execution Environment (TEE) for ARM processors specifically designed for IoT handheld systems. It provides memory isolation techniques to protect the trusted application data from being exploited by malicious entities.
% Certain studies show vulnerabilities where malicious entities may be able to gather private user information from the "secure world" by exploiting shared CPU resources like memory or cache. These high-performance shared memories were added by vendors to increase the processor's performance. 
In this work, we focus on identifying different vulnerabilities of the TrustZone extension of ARM Cortex-M processors. Then design and implement a threat model to execute those attacks. We have found that the TrustZone is vulnerable to buffer overflow based attacks. We have used this to create an attack called MOFlow and successfully leaked the data of another trusted app.
%\yuan{Consider naming your attacks differently, or people might think it is exactly the same as the original heartbleed attack.} 
This is done by intentionally overflowing the memory of one app to access the encrypted memory of other apps inside the secure world. We have also found that, by not validating the input parameters in the entry function, TrustZone has exposed a security weakness. We call this Achilles' heel and present an attack model showing how to exploit this weakness too. Our proposed novel attacks are implemented and successfully tested on two recent ARM Cortex-M processors available on the market (M23 and M33).
\end{abstract}


\section{Introduction}\label{sec:introduction}

 \begin{figure}[htbp]
\centering
\includegraphics[width=0.4 \textwidth]{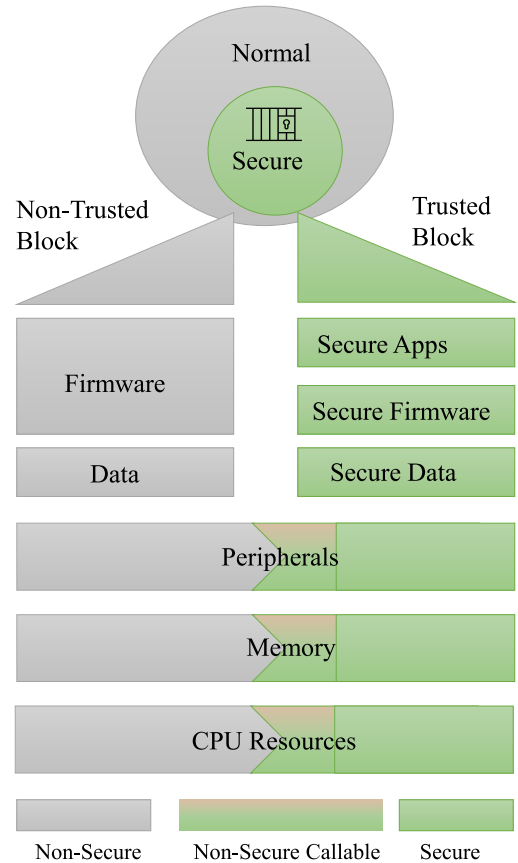}
\caption{TrustZone Core Virtualization}
\label{fig:TrustZone}
\end{figure}

ARM TrustZone is an embedded security system for ARM Cortex processors. Recently, ARM included TrustZone into IoT computing with cortex-m processors. 
%TrustZone starts at the hardware level, creating two distinct environments that run simultaneously (Figure \ref{fig:TrustZone}): the secure world and the normal world. 
The benefit of TrustZone is its compact and lightweight nature, allowing for both worlds (Figure \ref{fig:TrustZone}) to operate on a single processor core. Because of this secure operating system, ARM micro-controllers can store all system-essential libraries and applications in a secure area \cite{TrustZone1}. The defense mechanism in TrustZone is to protect memory (physical and cache) and process. For example, memory in both worlds is isolated with a security attribute Unit (SAU), even the same app with different signatures running in two different worlds has to go with a robust verification process and execute in isolation. Work stretching across different applications, both secure and not secure, can do so through a software-based secure monitor which mediates between the two security worlds. This software-based secure monitor is executed on the same core as all the other processes, and thus consumes less power than the traditional approaches detailed above. Even with this, there are malicious attacks by observing entry and exit onto the address of cache, compromising the messaging channel between a non-secure process and secure process\cite{liu2020cost}. To target this problem, some research papers are used isolated cache protection design to narrow down the access space. 
%All software in a TrustZone system operates in either the secure or normal world, never both. 
%ArupNote{Change it} \textcolor{blue}{\textbf {This secure system is referred to as a “Trusted Execution Environment”, or TEE.}} 

Due to limited or not availability of cache, access in memory inside Trusted Execution Environment(TEE) is shared and not bound to specific secure kernel process. So, the TEE delineates specific memory addresses in accordance with their world. There is no tightly coupled memory dedicated to a specific app in the secure zone. A trusted execution environment can be easily exploited by leaking memory within the shared space. Without authorization or access to protected enclaves, the attacks can be quite effective at collecting the users' private and secure data. This allows sensitive information to be stored out of reach for applications operating outside of the secure world. 

%\yuan{Up to now, most content you wrote seemed to be the background. Instead, you should write more about the uniqueness of trust computing on IoT, and if that provides opportunities or challenges for side=channel attacks.}

%\ArupNote{Re-write the below paragraph}

With the use of ARM TrustZone in the IoT ecosystem, memory access in these devices have a significant research focus within single and cloud with multiple connected smart devices.
%Because ARM TrustZone presents a new era in IoT security, 
The goal of our study is to research and develop security exploit encrypted information to gather sensitive user information into the normal world. 
%In other words, we intend to identify a vulnerability allowing encrypted information to leak in the secure world. 
%To do that, we have studied the architecture and design of the memory footprint of the TrustZone framework to find any potential vulnerability during the communication between secure and non-secure world.
Although the security attribute Unit (SAU) ensures the security, certain input parameters might expose the access of secure memory if the developer forgets to check memory-bound checking non-secure callable zone. The system should have an automatic guard to validate the memory-bound checking. Poor implementation at the nonsecure callable side might expose the potential loophole. This will create multiple openings for external attacks. A secure framework should not have APIs to get non-accessible data from the user level. TrustZone does not have any automatic internal memory management like with a high-level programming language. Moreover, security design and protection work differently in x86. Most of the low-level APIs are primitive and do not have any metrics to benchmark the security level. We have not found the API security validation from the ARM platform. A developer has to perform extensive operations for allocating memory and clearing them. Any intentional or unintentional memory leakage might expose the sensitive data even from the secure TrustZone memory.

The proposed threat model of MOFlow is based on the experimental results and found memory leaks during the access to out-of-bound data even in the TrustZone secure world. We also find, using invalid parameters in the Entry function, it is possible to infiltrate the secure world. We call this an Achilles’ heel for the TrustZone security. These successful attacks will highlight security vulnerabilities in the current ARM Cortex-M processors which need to be addressed to ensure the safety of the IoT systems. This will also help us understand potential risks associated with TrustZone and improve the security of IoT trust computing.

%\ArupNote{Rewrite it with the linking of specific contributions} 

In short our contributions in this work are:

\begin{itemize}
    \item We have done a robust exploration of the security vulnerabilities during the communication in between normal and secure world in the ARM TrustZone Cortext-M processor and defined open scopes of possible compromise of the system. 
    \item We propose a threat model to exploit memory overflow with intentional or unintentional fraudulent communication, encapsulated with security attribute unit along with mechanism for creating Achilles's heel. 
    \item We also expose the APIs limitations and the implication of a low-level framework that creates a possible loophole for the intruder.
    \item We provide best practices for the defense improvement inside TrustZone based on the experimental results and analysis that includes an additional layer of verification.
    \item Finally, we propose a trust model with TrustZone extension APIs and verifier along with communication flows.

\end{itemize} 

\textbf{Paper Organizations}. The rest of the paper is organized as follows. Section \ref{sec:background} presents the backgrounds on TrustZone and its architecture. Section \ref{sec:overview} explains the motivation behind the attacks and what was the expected outcome. Section \ref{sec:threatModel} has the design of the threat model. Then Section \ref{sec:apply} shows how we planned to apply it to ARM TrustZone. Section \ref{sec:experimentalSetup} presents the experimental setup. In Section \ref{sec:attacks} we list the different types of attacks we performed on the TrustZone and its results. Discussions on the implications of our findings, possible mitigation plans against the attacks and future works are added in Section \ref{sec:discussions}. Section \ref{sec:relatedWorks} lists the related works. Section \ref{sec:limitations} contains the limitations of our work. And finally, Section \ref{sec:conclusion} has the conclusion.

%\yuan{Even if this is going to be an attack paper, we would still need to discuss a bit about potential defenses.}

\section{Background}
\label{sec:background}
%\ArupNote{Re-write the below paragraph}
There are multi-variety of designs in ARM TrustZone to ensure security. ARM Cortex M23\cite{arm-m23} and M33\cite{arm-m33} do not have any in-built cache because of the compact design and priority on security features. 
%The initial focus of this project was identifying a possible vulnerability in the TrustZone-M architecture conceptually. In the beginning, without access to physical ARM TrustZone-M micro-controllers, the team took a special interest in cache-based attacks. 
In ARM Cortex-M35P, the process cache is the primary element of in-memory design to create a bridge between the processor execution and the relatively slower memory access. In the TrustZone-M design both instruction and data, a memory is expanded with an additional feature called an NS flag which helps to identify the security domain. This flag bit will be used to isolate the memory. These lines are not accessible from the normal world directly. But it is common for both worlds, during the execution of the processor. So the normal and secure world will try to use this memory line to support its running application.

%\yuan{I feel this story should not be included in the paper. Instead of telling your experiences directly, try to organize the content around the facts in the systems.}

 \begin{figure}[htbp]
\centering
\includegraphics[width=0.48 \textwidth]{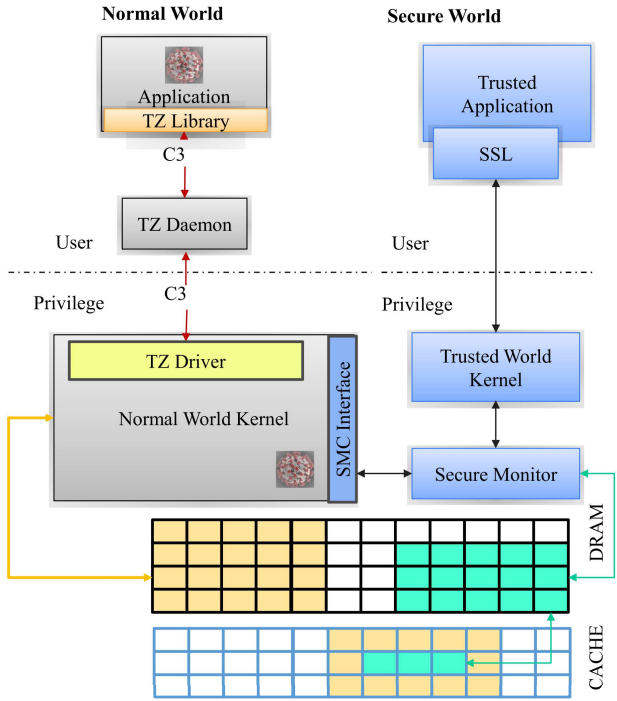}
\caption{A standard communication by using shared resources}
\label{fig:TrustZoneExecution}
\end{figure}

%\ArupNote{Re-write with the current state of the design}
The main reason for this design is to maximize the utilization of the memory and improve system performance. ARM sets specific hardware to secure the access of memory by any world application. But the access pattern is not secure in simple designed cortex M33 or M23 where only a single memory unit is available. Moreover, M55\cite{arm-m55} has a robust memory with instruction and data. These will communicate with customs-designed newly introduced instruction and data Tightly-Coupled Memory (TCM). Access patterns between TCM to cache can be easily monitored by an attacker process, leaving TrustZone vulnerable for the cache access side-channel attacks. From the beginning of trust computing, there is a vast number of studies on Intel-based SGX secure container\cite{yxu2015cache, shih2017cache}, but very few studies are done on TrustZone \cite{lipp2016execution}, running on mobile platforms. A graphic overview of the cache-based attack is seen in Figure \ref{fig:TrustZoneExecution}. 

%Unfortunately, due to the global chip shortage, the team was not able to attain an M55 or M35 cache-enabled ARM TrustZone micro-controller. For this reason, we will forego cache-based attacks moving forward and defer them to future work.

%Without access to cache-enabled ARM TrustZone-M micro-controllers, we are obligated to look elsewhere for security vulnerabilities. 
This security probing involved reading literature regarding the TrustZone-M architecture, control flow, and components. After re-framing the approach, the team began looking into previous effective cyber-attacks and the fundamental principles behind them. Though TrustZone-M provides a lot of new obstacles for attackers to overcome, we believed that certain attack models could be modified and applied to this security architecture. Throughout our literature exploration, we came across the MOFlow bug \cite{HeartBleed}.

%\yuan{Are you doing something exactly the same as the original attack? What are the differences? What are the novel challenges or opportunities you used for the attacks here?}

%\ArupNote{Remove heartbleed analogy from background}
%The HeartBleed Bug was initially discovered in 2014 in the OpenSSL cryptographic software library. OpenSSL was immensely popular at the time and used to secure communications on the internet using secure sockets layer and transport layer security encryption \cite{ssl}. OpenSSL is and was widely used on most HTTPS websites. When the HeartBleed bug was initially discovered, it was present on thousands of web servers, including major websites such as Yahoo. By exploiting this bug, malicious users could trick servers into sending over sensitive information such as usernames and passwords. 

The MOFlow bug relies on a TEE/Non-secure communication with the standard API of TrustZone. A nonsecure app sends a short message to a secure app to check if the app is active in the background. When a secure UI/service app falls out of responses with another due to inactivity or being killed or crashed, it is needed to be able to check if they are still alive. There will be data inconsistencies due to interaction by the user or server or any connected apps in the IoT system. That encrypted piece of data is sent from another node to check its status or availability. When the crashed or killed secure node receives this request, it responds with the same piece of encrypted data to prove to the nonsecure app that the secure app is still in place. This is where the vulnerability lies. The request message also includes information about its length. Below, we will draw out a scenario of how communication can be used to extract information from a secure app \cite{HeartBleed1}.

A normal world app is a malicious user and wants to extract sensitive information from the secure world app. So communication from Non-secure will go to secure for checking the availability of the service. The request consists of an encrypted message (e.g, 16KB lengths), but the normal world app intentionally lies about the length of the encrypted message and says that it is 128KB long (the maximum request length). The secure app receives this request and allocates a 128KB memory buffer to contain the encrypted message it is supposed to send back to the malicious app in Non-secure. The secure world then stores the 16KB encrypted message on the 128KB memory buffer and sends it back to the Non-secure. This is where the vulnerability lies. Security attributes in secure zone do not verify that the encrypted message length is equal to the length value provided. This tricks the secure world app into sending over 112KB of possibly sensitive information.

%\yuan{I feel lots of information in the background is actually helpful. You can consider moving some useful content to the approach section. That's because some reviewers might not read the background if they think they are familiar with the topic.}

The lack of this safeguard on secure zone allowed malicious users to use the MOFlow vulnerability to extract data from unsuspecting secure world apps. In this work, we intend to perform the MOFlow attack to target an ARM TrustZone-enabled micro-controller. 
%In the traditional HeartBleed attack, there is communication between the malicious computer and the targeted server. This communication manifests in the form of the OpenSSL software. In our application, we want to communicate between the TrustZone normal world and the secure world. More specifically, a malicious application on the TrustZone normal (non-secure) world is the malicious user (attacker), and sensitive information stored in the TrustZone secure world represents our target. Throughout this paper, we will detail the approach used to extract sensitive application information from the secure world.
This attack requires an important assumption which we will make for this experiment: the malicious user has planted a buggy application on the secure world. This is an important assumption because, without a buggy application within the secure world, there is no avenue for the attacker to interface with the secure world. While this obstacle seems difficult to overcome, we believe that it is a plausible scenario. With the onset of IoT systems, particularly smart homes, the user is free to download and use third-party applications that provide additional features. This app store provides an avenue through which a malicious user could plant a seemingly innocuous application that contains a bug enabling MOFlow attacks. Any user that downloads this malicious application opens the door for attackers to execute the MOFlow attack on a TrustZone-enabled device.
In Section \ref{sec:MOFlow}, we have discussed the attack model tailoring MOFlow to TrustZone-enabled micro-controllers.

\section{Overview of Approach}
\label{sec:overview}

\subsection{A Motivating Use Case}

To provide a motivating example, suppose there exists an IoT smart home device that is powered by a TrustZone-M enabled micro-controller. This smart home device can be connected to sensors such as a user’s smartwatch device, house lights, front door, and many other miscellaneous smart household IoT devices. The smart device can also interact with multiple cloud servers for each app that provides users the functionality to make purchases, check health statuses, and send messages and emails. If the MOFlow attack is proven to work within TrustZone-M devices it could lead to serious violations in the integrity of TrustZone’s security measures. Specifically for the described TrustZone-M powered IoT device, an attacker can publish a malicious application with memory leakage to the device’s affiliated marketplace and disguise the application as a seemingly innocent service that a user could end up downloading (similar to utterance checking) into their smart device’s secure zone. TrustZone applications can retrieve sensitive information from the server to get access to a sensor and save it to the memory. From there, the malicious application is among different other legitimate applications for sensors that could have the functionality to retrieve sensitive information from a using shared memory space. The attacker could then invoke this compromised secure-world application by overflowing the secure memory space. If another sensor’s data is saved on the device, the attacker could gather the user’s device identifiers, device authentication key, and other data from the memory.

\subsection{Expected Robustness Properties}
Let's define communication properties between \(Non-secure\) and \(TEE\) with a set of blocks instructions $X\{x_1, x_2, x_3, ... , x_n\} \Leftrightarrow Y\{y_1, y_2, y_3, ..., y_n\}$. If \(\triangle m\) is the leakage memory, then the response of \(X\) from the \(TEE\) is, 
\begin{align*}
    R_X &= O_Y + \triangle m
\end{align*}
where \(O_Y\), is the the expected allocated memory.

The model tries to perform the maximum number of attacks on \(TEE\) and increase the number of successful attacks \(S_N\). Target is to maximize the amount of leaked memory, \(F_m\) with the generator function L. So,
\begin{align*}
    \lim_{\triangle m\to\ F_m} f(\triangle m) &= L
\end{align*}

So for all instructions \(X\{x_1, x_2, x_3, ..., x_n\}\), output response is generated with multiple equations as follows,
\begin{align*}
    R_{x_1} &= {\triangle m}_{x_1} + O_{x_1} \\
    R_{x_2} &= {\triangle m}_{x_2} + O_{x_2} \\
    R_{x_3} &= {\triangle m}_{x_3} + O_{x_3} \\
    &\cdots \\
    R_{x_n} &= {\triangle m}_{x_n} + O_{x_n} 
\end{align*}

To verify the robustness properties of \(TEE\) secure communications, \(\triangle m\) should be \(0\), e.g., 
\begin{align}
    \triangle m &= {\triangle m}_{x_1} + {\triangle m}_{x_2} + {\triangle m}_{x_3} + ... + {\triangle m}_{x_n} = 0
\end{align}
In this paper, by performing a set of attacks, we will invalidate the robustness properties of \(TEE\).

\subsection{Aligning Problems on ARM TrustZone} 
 We have done a robust study on normal-world user and kernel space and have learned of vulnerabilities allowing attackers to gain full control of the normal-world kernel space. It is possible to discern physical addresses from virtual information. Address translations play a vital role in allocating memory and are thus a prime area for an attack. By design, the whole memory is divided into multiple parts. Our first target is to find a path to access the secure memory. Moreover, the cycle counter can be used as a precision timer that is accessed by only super users. In addition, a non-privileged app can access information without super-user permissions and with no virtual to physical address translation or cycle count. 
 
 This creates an opportunity for prime and probe attacks.  To do that, there can be multiple scenarios. When a normal world app tries to access securely by not following the standard protocol, on the framework side, there should be some security measures to protect any kind of illegal access. Security Attribute Unit(SAU) and Implementation Defined Attribution Unit(IDAU) will raise kernel fault in response. What if the developer made the mistake of adding memory boundary checking in the non-secure callable? A normal world app will have access to the whole memory of a secure world. Many high-level programming languages have inbuilt garbage collectors to free allocated memory and handle memory leaks. If a system does not have a built-in garbage collector, it should have support at the framework level to handle memory leakage internally. 
 
 ARM TrustZone is based on low-level language, Assembly, and C. In these languages, developers have to manage every allocated memory checking. One of the major limitations in the ARM TrustZone framework is, it does not have any in-built memory management support, even for secure zones. This opens the door for the overflow of the memory in a secure zone and possible leakage of valuable data. 
 In Figure \ref{fig:TrustZoneExecution}, communication line C3 is the main way between normal world user and kernel space. With C3 superuser access, a non-privileged app gets access information without cycle count and address translation. C3 is executed with a TrustZone daemon or library which needs an extensive authentication process for the execution in a secure world. But C3 has access to a nonsecure callable. Intentional memory accessible is possible with bad coding and generates Achilles' heels. An attacker can get overflow memory data by using standard TrustZone API. No other apps, including TrustZone itself, will have a single idea about the theft of the information.

\subsection{Threat Model Design}
\label{sec:threatModel}

Based on the design by ARM, all cryptographic operations are executed in an isolated environment \cite{zhang2016case, zhou2014case}. That means API execution in a process of a cryptographic library like SSL is isolated in the secure world. We have designed our threat model based on the assumptions that there must be a channel of handshaking between the normal world and secure world data or instruction transmission. If those operations happen either on the SMC interface or TZ manager, then the attacker can easily get data by using standard protocol from a secure world and extracting necessary information to get the AES key. Because Zhang et al. \cite{zhang2018trusense} demonstrates a way of recovering the full AES128 key using the application level attack in a shorter time.

Now the main idea is to get data from the memory by overflowing the assigned data structure. All apps in the secure world use shared resources. Assigning memory to an app is a loosely coupled operation at the processor. If a malicious app overflows its memory scope, it can easily get data that was not assigned. Although the data is encrypted, it can be easily decrypted by using a T-table-based decryption mechanism. Moreover, input parameters play an important part in getting the level of access to a secure world. There is no standard system in the TrustZone framework to handle any fuzzy attacks. Developers might not check all the corner cases of access memory in non-secure callable parts. SAU and IDAU do not guarantee parameter level verification at non-secure callable regions. Here comes the Achilles heel. With that attacker can compromise the non-secure callable and get full access to secure world memory.

The proposed threat model will work from the application level with user privilege, which does not have any assumption to break the hardware-enabled trust execution environment. So executing the code from normal world user space to kernel space does not need any API call or permission from the TZ library or TZ manager in kernel space. A malicious process in a secure space can run and infect any operation and remain intact inside an app. This process might have access to memory data with the backdoor leakage. Based on this analogy, this threat model is more resilient in the IoT system and does not need any dependencies on the TrustZone specific platform. Based on this threat model, suppose, an attacker has both a secure and non-secure app, running on an IoT device, and he wants to steal information from other vendors' apps running on the same device. 

\begin{figure}[htbp]
\centering
\includegraphics[width=0.48\textwidth]{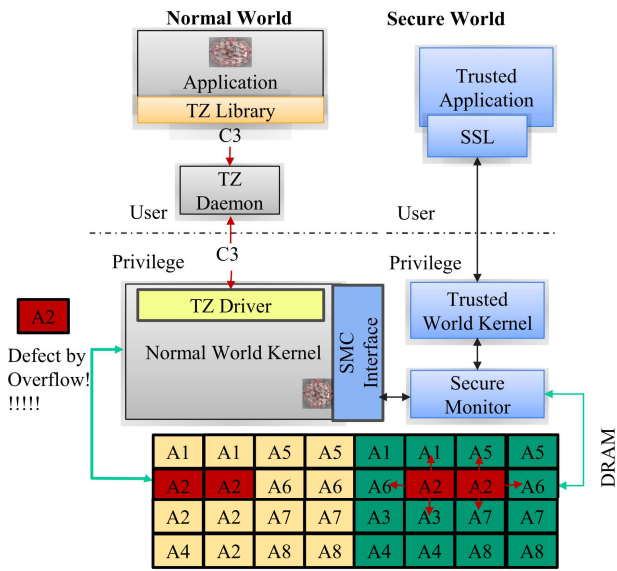}
\caption{The Proposed Threat Model}
\label{fig:TrustZonethreatmodel}
\end{figure}

In Figure \ref{fig:TrustZonethreatmodel}, A2 is the malicious app that memory leaks. In ARM TrustZone, there is no support for handling malicious memory overflow, inside a secure zone. So, A2 will read data from the DRAM which was assigned to any other app, and send it back to the normal world by following the APIs of non-secure callable. Because in TrustZone memory, there is no tightly coupled memory bound to a specific app. As a result, even the TrustZone framework and no other app will detect the theft of information. For the simplicity of the threat model, we have excluded the decryption mechanism of secure data from the project scopes. 

\begin{figure}[htbp]
\centering
\includegraphics[width=0.48\textwidth]{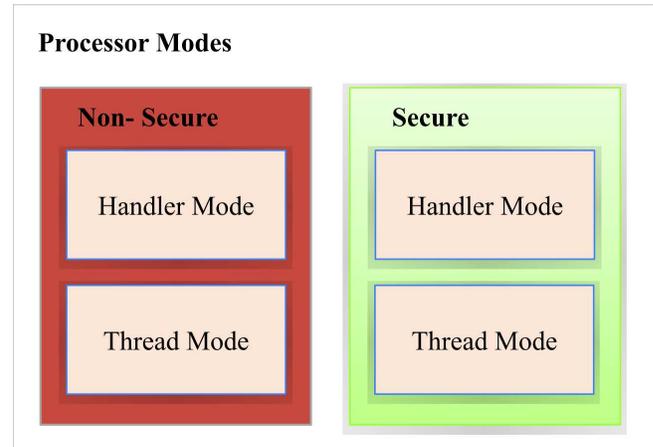}
\caption{ARM Cortext-M micro-controller modes}
\label{fig:processormodes}
\end{figure}

\subsection{Apply Threat Model to ARM TrustZone}
\label{sec:apply}
%\ArupNote{Discuss the architecture security attributes}
ARM Cortext-m is designed as a component of IoT ecosystem. As it is low power, TrustZone security extension is optional. That means, chipset vendor has the flexibity to design chip. For example, NXPLPC55S28 is based on Cortex-M33, but this board does not have TrustZone security extension. As it is low powered micro-controller, processor works differently than ARM cortex-A. ARM Cortex-M processor works in two different modes in Figure-\ref{fig:processormodes}. When running application software, the CPU is in Thread mode, and for handling exceptions, it is in Handler mode. When the processor exits reset, it enters Thread mode and exits Thread mode when all exceptions have been processed. Execution can be privileged or unprivileged in Thread mode. Execution is Privileged in Handler mode. Memory maps are used to divide the Secure and Normal worlds, and transitions are handled automatically in exception handling routines.That's why multiple secure function entry points are supported by Armv8-M \cite{armtrustcortexm2021}.

Because of that, all access to different memory might be on multiple in parallel. Although SAU and IDAU protect the memory access with NS bit, what is transmitting from the secure zone does not have any control. Moreover, in both thread and handler mode, within the region of secure memory, data access is performed based on the programming logic of secure memory. Attribute units are independent and do not have any influence on application features. This design opens research questions about the security flaws inside secure and non-secure callable and that's how our proposed thread model has implications on the secure zone.

 \section{Experimental Setup}
 \label{sec:experimentalSetup}

Multiple vendors develop  board based on ARM Cortex-M along with development environment. Our primary analysis for feasibility test, was started with QEMU emulator for RPI3 kernel in linux \cite{qemu2022, raspberrypi2021, rpilinux2021}. But we were unable to replicate the defined problem in target domain. Because, it doesn't have TrustZone framework and the architecture is not comply the current state of the arts. NXP and Nuvoton released R\&D board based on ARM cortex-M and we have used NXPLPC55S69 \cite{nxp55s69} and \cite{nuvoton2351} Nuvoton M2351. Nuvoton-M2351 has a single core M23 processor and NXPLPC55S69 has a dual-core M33 processor with DRAM. Both of them have support for TrustZone instructions. Our initial plan was to use the Cortex-M35P and M55 processors. Because they have the latest TrustZone implementation. Cortex-M55 has additional instruction and data tightly coupled memory. These are configurable to the specific app for the fixed memory location. Unfortunately, we couldn't get either of them publicly available on the market. Or even if they were available, there were substantial amount of time delay for the delivery due to chip shortage. So, we chose the M23 and M33-based boards.
% in Figure \ref{fig:experimentalboard}. 
We have also received an NXPLPC55S28 board, developed with a single ARM cortex M33 processor. But it does not have any support of TrustZone, so it couldn't be used in this work. 

%\ArupNote{Remove below lines} Because of not supporting cache memory in those boards, it narrowed down the overall design scope to in-memory operation. 

% \begin{figure}[htbp]
% \centering
% \includegraphics[width=0.45 \textwidth]{Figure/experimentalboard.pdf}
% \caption{Experimental boards for the threat model}
% \label{fig:experimentalboard}
% \end{figure}

\section{Attacks on ARM Cortex-M}
\label{sec:attacks}

We have performed multiple attacks on Cortex M processors. Some attacks are failed due to security properties by ARM. Failed attacks are an Invalid transition from secure to the normal world, the invalid entry point from normal to secure world, and invalid data access from the normal world. We do have some success. Success attacks are Invalid input parameters in the entry function, we call it Achilles' heel and steal Memory data inside a secure world, we call it Heart Bleed. In the next subsections, we will describe in detail all attacks. Source codes for all of the attacks are publicly available on
\textbf{\textit{\textcolor{blue}{\url{https://github.com/arupcsedu/MVAM}}}}.

\subsection{Memory Map}
\begin{figure}[htbp]
\centering
\includegraphics[width=0.48 \textwidth]{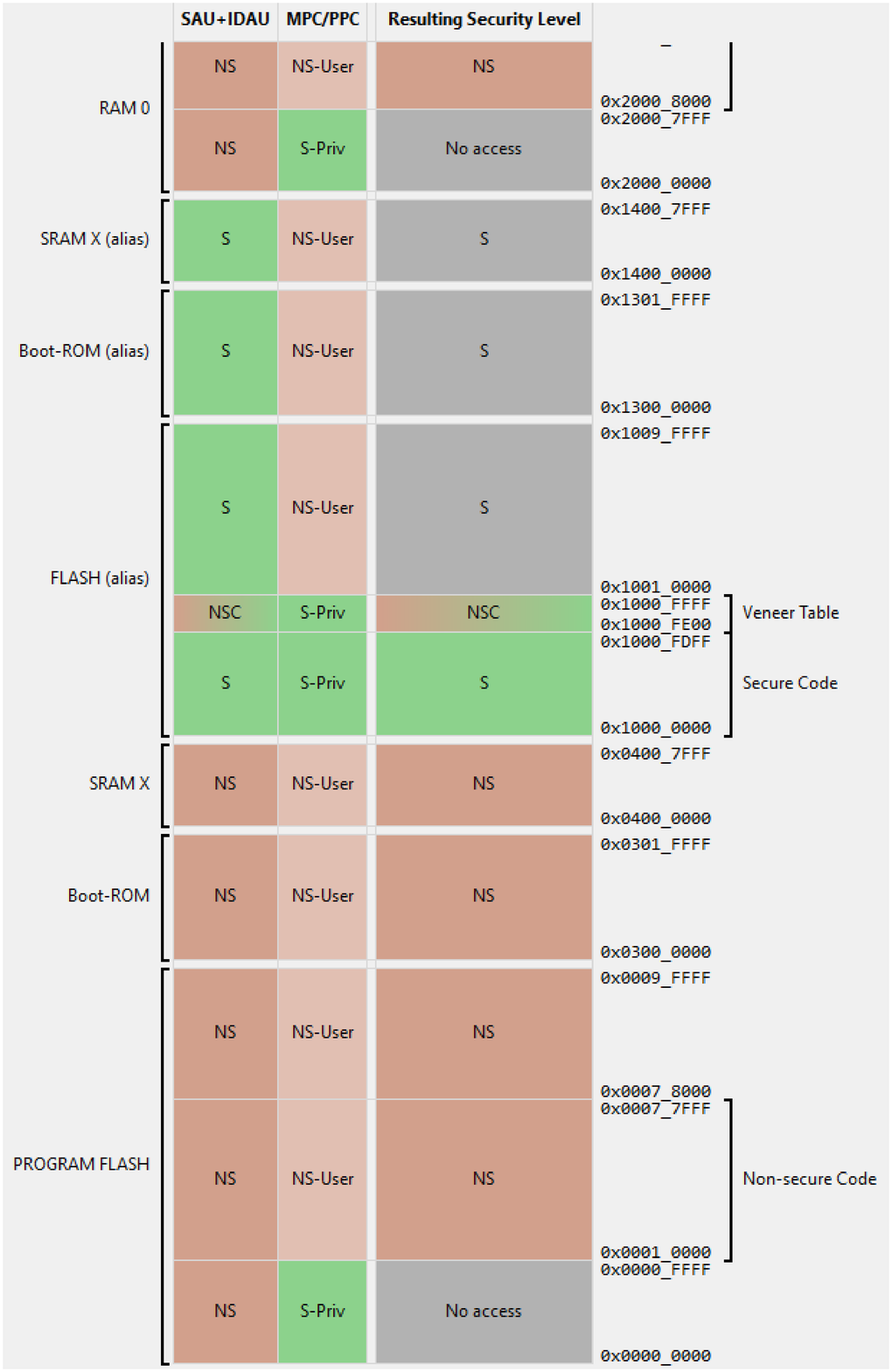}
\caption{Memory map of Secure, Non-Secure and Non-Secure Callable}
\label{fig:m33memorymap}
\end{figure}

Before going into details about our experiments, let’s check the run-time memory attribute map of ARM Cortex-M in Figure \ref{fig:m33memorymap}. We have exported this memory snapshot from the LPCNXP55S69 board, during running the program. We see the NS Program flash base is 0x0001\_0000. The Secure Program flash base is 0x1000\_0000. A Non-secure Callable, here with NXP, we call a Veneer Table, the entry point to secure area base is 0x1000\_FE000. A combination of SAU (Secure Attribute unit) and IDAU (Implementation Defined Attribution Unit) ensures the separation of each memory footprint with security. Here SAU is internal with a processor and IDAU is external units, normally designed by chipset vendors, for example, NXP has that flexibility to design IDAU.

\subsection{Invalid Transition From Secure to Normal World}
In this attack, a direct address to non-secure RESET is used to jump into the normal world. There are two issues related to this approach in Listing-1. \textbf{First}, all core registers are not clear so there is a potential data leak. \textbf{Second}, the most LSB of address into the normal world has to be cleared. We have not performed those and the requirement is not met for the transition to the normal world. As a result, a secure fault is generated by SAU.

%\begin{figure}[htbp]
%\centering
%\includegraphics[width=0.45 \textwidth]{Figure/ITSecureToNonsecure.pdf}
%\caption{Attack with Invalid Transition From Secure to Normal World }
%\label{fig:ITSecureToNonsecure}
%\end{figure}

\begin{lstlisting}[caption=Attack with Invalid Transition From Secure to Normal World]
#define CODE_START_NS 0x00010000
#define AHB_LAYERS_COUNT   12U
#define NON_SECURE_START CODE_START_NS

if (testCaseNumber == FAULT_INV_S_TO_NS_TRANS)
{
    funcptr_ns ResetHandler_ns;
    /* Non-secure main stack address */
    __TZ_set_MSP_NS(*((uint32_t *)(NON_SECURE_START)));

    /* Initialize the non-secure vector table */
    SCB_NS->VTOR = NON_SECURE_START;

    /* Function pointer for the Non-secure reset handler */
    ResetHandler_ns = (funcptr_ns)(*((uint32_t *)((NON_SECURE_START) + 4U)));

    /* Invalid switch to non secure */
    __asm("BXNS %0" : : "r"(ResetHandler_ns));
}
\end{lstlisting}

Both issues can be solved by using the \_\_cmse\_nonsecure\_call keyword attribute. If this attribute is used for a function call to a normal world, the compiler will do three things. \textbf{First}, clear all used registers to avoid potential data leak. \textbf{Second}, clear LSB address bit. \textbf{Third}, jump to address using BXNS instruction. The BXNS instruction causes a branch to an address and instruction set specified by a register and causes a transition from the Secure to the Non-secure domain. This variant of the instruction must only be used when additional steps required to make such a transition safe are taken \cite{bsnxins}.

\subsection{Invalid Entry From Normal to Secure World}
In Listing-2, a function pointer, PRINTF\_NSE is intentionally increased by 4. It is defined with a non-secure callable function DbgConsole\_Printf\_NSE in the veneer table. By this the Secure Gateway(SG) instruction is skipped, when a function is called. This causes an illegal entry point into a secure world and a secure fault is generated. The correct entry point into the secure world is ensured by using \_\_cmse\_nonsecure\_entry keyword attribute for every entry function so that it clears the register value and LSB address bit. Then the linker creates a veneer table for all entry functions with SG instructions.

%\begin{figure}[htbp]
%\centering
%\includegraphics[width=0.45 \textwidth]{Figure/ITNonSecureToSecure.pdf}
%\caption{Attack with Invalid Transition From Normal to Secure World }
%\label{fig:ITNonSecureToSecure}
%\end{figure}

\begin{lstlisting}[caption=Attack with Invalid Transition From REE to TEE]
#define SEC_ADDRESS    0x10000000
#define NONSEC_ADDRESS 0x20130000
typedef void (*funcptr_t)(char const *s);
#define PRINTF_NSE DbgConsole_Printf_NSE


if (testCaseNumber == FAULT_INV_S_ENTRY)
{
    func_ptr = (funcptr_t)((uint32_t)&PRINTF_NSE + 4);
    func_ptr("Invalid Test Case\r\n");
}

/* Non-secure callable (entry) function */
TZM_IS_NOSECURE_ENTRY void DbgConsole_Printf_NSE(char const *s)
{
    size_t string_length;
    /* Access to non-secure memory from secure world has to be properly validated */
    /* Check whether string is properly terminated */
    string_length = strnlen(s, MAX_STRING_LENGTH);
    if ((string_length == MAX_STRING_LENGTH) && (s[string_length] != '\0'))
    {
        PRINTF("Input data error: String too long or invalid string termination!\r\n");
        abort();
    }

    PRINTF(s);
}
\end{lstlisting}

\subsection{Invalid Data Access From Normal World}

In Listing - 3, the pointer is set to the address defined by NONSEC\_ADDRESS. This address has a non-secure attribute in SAU but it has a secure attribute in AHB secure controller. If data is read from this address, the data bus error is generated. Compared to attacks for accessing the memory address, SEC\_ADDRESS where the secure fault is generated, this error is caught by AHB secure controller, not by SAU. Because in the SAU this address is non-secure. So the access from the normal world is correct from SAU's perspective. In the normal world, the application does not have access to secure memory.

%\begin{figure}[htbp]
%\centering
%\includegraphics[width=0.45 \textwidth]{Figure/IDAFroNormaltoSecure.pdf}
%\caption{Attack with Invalid data access to Secure}
%\label{fig:IDAFroNormaltoSecure}
%\end{figure}

\begin{lstlisting}[caption=Attack with Invalid data access to TEE]
#define SEC_ADDRESS    0x10000000
#define NONSEC_ADDRESS 0x20130000
typedef void (*funcptr_t)(char const *s);
#define PRINTF_NSE DbgConsole_Printf_NSE

if (testCaseNumber == FAULT_INV_NS_DATA_ACCESS)
{
    test_ptr   = (uint32_t *)(SEC_ADDRESS);
    test_value = *test_ptr;
}
\end{lstlisting}

\subsection{Achilles' Heel - Invalid Parameters in Entry Function}\label{sec:achillesHeel}

In this attack, the input parameter is set to address \textbf{0x10000000} in Listing-4. This address has a secure attribute (see SAU settings in the memory map picture). This secure violation is not detected by secure fault, since the input parameter is used by the secure function in a secure mode. So this function has access to the whole memory. However, every entry function should check the source of all input data to avoid potential data leaks from secure memory. The correctness of input data cannot be checked automatically. So, this function is an Achilles' heel, which can be used to enter the secure world by using a valid secure location as an input parameter. This has to be checked by software, using TT instruction by publisher vendors to protect Achilles' heel if the developer forgot to set a check in the NSC layer.

%\begin{figure}[htbp]
%\centering
%\includegraphics[width=0.45 \textwidth]{Figure/AchillesHeel.pdf}
%\caption{\ArupNote{Find better ways to write code block}Achilles' Heel Attack during access Secure World with invalid input parameters}
%\label{fig:achillesheal}
%\end{figure}

\begin{lstlisting}[caption=Achilles' Heel Attack during access TEE with invalid input parameters]
#define SEC_ADDRESS    0x10000000
#define NONSEC_ADDRESS 0x20130000
typedef void (*funcptr_t)(char const *s);
#define PRINTF_NSE DbgConsole_Printf_NSE


if (testCaseNumber == FAULT_INV_S_ENTRY)
{
    func_ptr = (funcptr_t)((uint32_t)&PRINTF_NSE + 4);
    func_ptr("Invalid Test Case\r\n");
}

/* Non-secure callable (entry) function */
TZM_IS_NOSECURE_ENTRY void DbgConsole_Printf_NSE(char const *s)
{
    size_t string_length;
    /* Access to non-secure memory from secure world has to be properly validated */
    /* Check whether string is properly terminated */
    string_length = strnlen(s, MAX_STRING_LENGTH);
    if ((string_length == MAX_STRING_LENGTH) && (s[string_length] != '\0'))
    {
        PRINTF("Input data error: String too long or invalid string termination!\r\n");
        abort();
    }

    /* Check whether string is located in non-secure memory */
#if (__GNUC__ != 10)
    if (cmse_check_address_range((void *)s, string_length, CMSE_NONSECURE | CMSE_MPU_READ) == NULL)
    {
        PRINTF("Achilles' Heel exception: String is not located in normal world!\r\n");
        abort();
    }
#endif
    PRINTF(s);
}
\end{lstlisting}

\subsection{MOFlow - Steal Memory Data Inside Secure World}
\label{sec:MOFlow}

Along with the Achilles' heel, we have implemented our threat model, MOFlow. In the MOFlow attacks, mentioned in Listing-5 a secure attacker app(A2) with memory overflow is running on the secure zone. Here, moflow() function is implemented in the secure app which memory leaks. There are three other test apps(A1/A3/A5) running on the TrustZone memory which does not have any leakage. Because of memory overflow in A2, it is getting more encrypted unassigned data from the memory which is allocated to other apps. A2 returns all data to the normal world by following the proper standard of TrustZone. Application, A1/A3/A5 and even TrustZone itself does not have a single idea about this stealing, as it is happening in a specific program space. With a T-table-based mechanism, it can be decrypted to actual data. Like in the MOFlow attacks, a secure zone is acting as a server and returning sensitive information to the normal world.

%\begin{figure}[htbp]
%\centering
%\includegraphics[width=0.45 \textwidth]{Figure/HeartBleed.pdf}
%\caption{MOFlow Attack on Secure World with buffer overflow}
%\label{fig:HeartBleed}
%\end{figure}

\begin{lstlisting}[caption=MOFlow Attack on TEE with buffer overflow]

#define FAULT_HEART_BLEED         0
#define FAULT_INV_S_TO_NS_TRANS   1
#define FAULT_INV_S_ENTRY         2
#define FAULT_INV_NS_DATA_ACCESS  3
#define FAULT_INV_INPUT_PARAMS    4
#define FAULT_INV_NS_DATA2_ACCESS 5
#define MAX_SMEM_SIZE 4e+9

TZM_IS_NOSECURE_ENTRY char* GetDRAMData_NSE(void)
{
	char leakData[MAX_SMEM_SIZE];
	char *lDataPtr = GetDRAMData();
	PRINTF("Read from Veneer Table:\n");
	for(int i = 0; i < MAX_SMEM_SIZE; i++)
	{
		leakData[i] = lDataPtr[i];
		printf("%c",leakData[i]);
	}
	leakData[MAX_SMEM_SIZE] = '\0';

	//strcpy(leakData, lDataPtr);
    return leakData;
}

char* GetDRAMData()
{
    return leakedData;
}

void moflow()
{
    char str[] = "I am malicious. Check my tail";

    testCaseNumber = FAULT_HEART_BLEED;

    int len = strlen(str);

    for(int i = 0; i < len + COM_DRAM_OFFSET; i++)
    {
    	leakedData[i] = str[i];
    	PRINTF("%c", leakedData[i]);
    }
    leakedData[len + COM_DRAM_OFFSET] = '\0';

    PRINTF("\nDecrypt the above data from my tail.\n");
    return ;
}
\end{lstlisting}

\section{Discussions}
\label{sec:discussions}

\subsection{Implications of Our Findings}

The Achilles' heel attack (Section \ref{sec:achillesHeel}) indicates it is important to check the memory locations as an input parameter. Without properly validating the inputs, they can be modified by an attacker and be used to compromise the execution of the target function. However, to the best of our knowledge, there is no automated tool available to detect invalid parameters. So developers would need to ensure their methods properly validate input parameters before using them for any sensitive process. Also, vendors would have to ensure software using their platform can prevent this kind of attack.
The MOFlow attack ( Section \ref{sec:MOFlow} ) takes advantage of the lack of tight coupling memory with applications that are using them. So any trusted application can access the memory of another trusted app and read the encrypted data. Even though the application data is encrypted, hackers may exploit the encryption algorithm used in ARM TrustZone to decrypt the extracted data. Lapid et al. \cite{lapid2018cache} showed using GPU-based analysis it is possible to crack the TrustZone implementation of AES. However, the SAU can be used to limit the applications from accessing others' data and thus resolve this vulnerability.

\subsection{Mitigation plan}
By design, TrustZone ensures the security to access the secure world. No unauthorized app can access any user or kernel service inside a secure world. But ensuring the security of data within TEE is challenging. Ron et al. \cite{stajnrod2021attacking} showed how an attacker can run arbitrary code in a secure world and how to handle those attacks with protection measurements. These are designed on top of control-flow attacks \cite{zhang2013control, davi2014hardware}. We will focus on the mitigation plan of protecting memory leakages and vulnerable points in non-secure callable so that any bad coding or intentional attacks are handled within the TrustZone framework. This will ensure the robustness of the system.

\textbf{Non-secure callable} give the bridge to a normal world app for sending any data or instruction to a secure world. Without the proper, guard for checking memory boundary in the veneer table, a potential Achilles heel will be created and compromised the whole system. There should be a mechanism inside the non-secure callable to check the memory boundary of a secure world. For example, in the Listing-6 below, \(cmse\_check\_address\_range()\) provides validity of incoming requests address range and blocks inside the non-secure callable regions for an Achilles heel.

\begin{lstlisting}[caption= Checking for a potential Achilles' Heel attack]
/* Check whether string is located in non-secure memory */
#if (__GNUC__ != 10)
    if (cmse_check_address_range((void *)s, string_length, CMSE_NONSECURE | CMSE_MPU_READ) == NULL)
    {
        PRINTF("Achilles' Heel exception: String is not located in normal world!\r\n");
        abort();
    }
#endif
\end{lstlisting}

A commercial application in the robust IoT ecosystems, multiple vendors will develop different kinds of services. To relay this kind of security checking on 3rd party application developers instead of automatic platform support is a risky design.

\textbf{The primitive APIs} for memory management are exposed by TrustZone. Process and executing the business logic of certain services is vital and error prune even for ARM platform developers. Furthermore, if the vendor application developer does not have an in-depth understanding of the underlying security design, the internal memory map can be messed up. As a result, attackers may be able to read sensitive data from other memory locations or trigger a system crash. When the code reads a variable quantity of data and assumes that a sentinel, such as a NULL in a string, exists to terminate the read operation, a crash can occur. 

If the expected sentinel isn't found in the out-of-bounds memory, too much data is read, resulting in a segmentation fault or a buffer overflow. Any instruction can change an index or execute pointer arithmetic on a memory address that is outside the buffer's limits. Following that, a read operation yields undefined or unexpected results.

To handle this, we are proposing an additional layer of security in between non-secure callable and secure zone. The purpose is to handle the abnormality of bad code inside the secure zone. This is expected that a 3rd party developer can write vulnerable code. The system should have a defense mechanism to find in various stages of development. We have not found any extensive tools to detect issues inside the secure applications with MCUExpress tools \cite{nxp-memory-profile2022} by NXP. ARM provides tools for memory profiling for other chipset \cite{arm-memory-leak2022}, not which has embedded TrustZone framework for 3rd party vendors. There are 3rd party C-based memory profilers \cite{mtuner-022, Valgrind-022, android-memory-profile2021} to analyze memory usage and highlight potential memory leak issues. But these are not customized for the profiling memory with security constraints. For example, root routes of new instances that could cause memory leaks. The root pathways provide information on why the instance is not freed. When determining how a memory leak happens, this is the most crucial information.

%\ArupNote{Add details about proposed framework architecture}

\begin{figure}[htbp]
\centering
\includegraphics[width=0.48\textwidth]{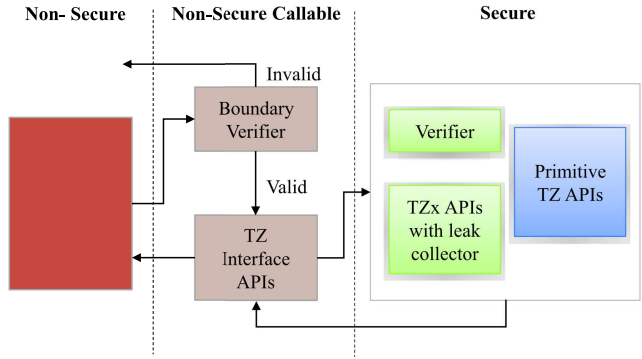}
\caption{The Proposed Trust Model}
\label{fig:TrustZoneModel}
\end{figure}

\textbf{To overcome MOFLow} attacks, we have proposed a communication design flow mentioned in the Figure-\ref{fig:TrustZoneModel}. There will be multiple components inside the non-secure callable and secure region of TrustZone and will comply with the security principle of SAU and IDAU.  

Non-secure callable (NSC) is a shared region (\ref{fig:m33memorymap}) for both secure and non-secure execution. \textbf{Boundary Verifier} will cross-check the request from non-secure instruction (e.g. \(x_1\)) and validate the address range. If it is valid, then \textbf{TZ Interface APIs} will be used for accessing the memory. Otherwise, an error (e.g. \(e_1\)) will be generated for the non-secure process. Non-secure app might be multi-threaded and executes parallel instructions. The purpose of the handling error in NSC to isolate defected instructions outside of the secure zone with meaningful information in error set (e.g. \(E\{e_1,e_2,e_3,...,e_n\}\)). This will block all possible Achilles' heels in NSC. 

We propose two additional components for the secure zone. \textbf{TZx APIs with leak collector} is an extension framework, consists of an API set, build on top of primitive TrustZone APIs. These APIs will have leak collectors in related logic and solve the problem of circular dependencies on the shared resource. \textbf{Verifier} in the secure zone plays a vital role to protect MOFlow attacks. Whenever a non-secure process will try to access any information which is held by a data structure, the verifier checks the boundary of the allocated memory before responding to non-secure process. Data security in shared memory with blocked Achilles's heels provides additional attributes for the robustness of the TrustZone.   

\subsection{Future Works}
It is possible to further extend our attack model by reducing the search space inside the TrustZone for the target data. This can help to trace the data for a particular trusted application inside the secure zone faster. Thus stealing the data only for that target application. For example, Chen et al. \cite{chen2017downgrade} proposed a cache flushing-based side-channel attack on the ARM processors to reduce the search space to find a specific key value within the cache memory.

\section{Related Works} 
\label{sec:relatedWorks}

Many of the recent TrustZone vulnerabilities are caused by cache attacks \cite{mushtaq2020winter}. Cache-based side-channel attacks mainly focus on the execution time and trace of user's accesses during the cache operations to perform these attacks. Lipp et al. \cite{lipp2016armageddon} used the lack of `cache flush` on old ARM cores (before ARMv8) to monitor cache activity within the ARM TrustZone from outside. The cache coherence protocol allowed processors to fetch shared cache lines and thus exposing them to cache-based attacks. Taking advantage of the coherence protocol in a multiprocessor system, Yarom et al. \cite{yarom2014flush+} was able to examine cache lines of one core from another by flush and reload attack. Lapid et al. \cite{lapid2018cache} exploited the misaligned T-table of the Keymaster Trustlet of ARM TrustZone in Samsung mobile and successfully extracted the AES-256 keys.

% Side-channel attacks have also been extensively used of these type of systems. Existing side-channel attacks can be categorized into several classes following which the observation method was used.

% \textbf{Power Analysis:} Use time-dependent differences in the device's power consumption during the procedures under observation. Recently machine learning techniques have been widely used to learn the correlations between the power usage and the target instruction \cite{hettwer2020applications}. \textbf{Electromagnetic Analysis:} Uses the electromagnetic radiation signatures produced by the target procedure \cite{leignac2019comparison}. \textbf{Acoustic Cryptoanalysis:} Exploits the properties of sound waves, e.g. the emission of certain frequencies and their change over time \cite{genkin2014rsa}. \textbf{Temperature Analysis:} Uses target's change of temperature over time when executing the sensitive operations \cite{hutter2013temperature}. 

Side-channel-based attacks also have been extensively studied on the ARM TrustZone. Chen et al. \cite{chen2017downgrade} was able to exploit a downgrade attack on TAs (Trusted Applications), by patching the old version onto the new one. The system's vulnerability would let others replace the current trust with an old vulnerable one and use that to run the TA. 

DMA (Direct Memory Access) attacks are also continuously under research. Yahuda et al. \cite{yehuda2020protection} showed that by dumping memory frequently using DMA transactions, write patterns can be examined. In ARM TrustZone, they were able to extract RSA keys. The DAGGER tool \cite{irazoqui2016cross} can steal cryptographic keys using a DMA-based keystroke logger. It can also attack the OS-kernel structure and file cache. 

The ARM debugging feature lets a host get read/write access to the TrustZone \cite{ning2019understanding} and leak private keys. The defective ECDSA signing in Qualcomm’s implementation of Android’s hardware-backed Keystore let attackers extract a 256-bit private key from the key store \cite{ryan2019hardware}.

\textbf{Current Work's Limitations:} Most of the attacks on ARM TrustZone focus on Cortex-A processors. However, the ARM Cortex-M processor is increasingly becoming more popular in Mobile and IoT applications. Because it is optimized specifically for them. Its design structure (fast hardware-based transition, no memory management, no-cache) is also much different from that of Cortex-A. So it is important to properly investigate possible vulnerabilities in its security protocols and TrustZone implementation.

\section{Limitations of Our Work} 
\label{sec:limitations}

The proposed attacks are done based on the assumption that we can install our vulnerable trusted application on the victim's device. This might not be possible in some cases where the attacker doesn't have access to the victim's device. However, it is possible to modify applications that the victim trusts and use that to install the modified vulnerable app. 

Even though our attack model has successfully extracted other applications' data from the secured zone, they are encrypted. So a separate tool will be needed to decrypt the data and make meaning out of it. However, some prior works have already been successful in cracking the encryption implementation of ARM TrustZone \cite{lapid2018cache}. So it is possible to overcome this limitation.
%\ArupNote{Re-write with new analogy of below paragraph}
Our used processor ARM Cortex-M33 is not the latest release with the ARM TrustZone feature. Despite our best efforts, we were unable to find any development boards in the market with the latex ARM Cortex-M35P and M55 processors. So the attack models might not represent an exact evaluation of the state-of-the-art ARM architecture and countermeasures. However, due to the short time limit of the project, it was impossible to wait for development boards with a very long delivery time.

\section{Conclusion}
\label{sec:conclusion}

After performing a series of different attacks on the ARM Cortex-M micro-controller with the proposed threat model, the MOFlow and Achilles heel approaches were successfully able to access encrypted data from the secure world region. However, there are some key limitations and controlled factors that make this vulnerability less likely to occur organically. The successful MOFlow attack can only be performed if the attacker can gain access to the secure world of a TrustZone’s secure region. A potential route of work to improve the likelihood of a successful MOFlow attack in the wild is finding a way to reduce the search space in the secure zone region. Secondly, retrieving the victim’s sensitive data from the TrustZone-M micro-controller is only one step in the process. Since the information is encrypted in the secure region, an attacker would need to exploit the correct decryption algorithm that TrustZone uses to obtain the plain-text information. One future route of work would be to investigate the implementation of TrustZone-M’s encryption and decryption algorithms and try to exploit them from the micro-controller. Doing so would enhance our current work significantly. Another opportunity for future work would be performing CacheTrack side-channel attacks on the Cortex-M35P or Cortex-M55 micro-controllers once their demand in the micro-controller market decreases. The Cortex-M35P and Cortex- M55 processors are considered state-of-the-art chips for TrustZone-M computing with instruction and data tightly coupled memory and there is a lack of research exploring these specific chips for novel vulnerabilities.

\bibliographystyle{IEEEtran}
\bibliography{main.bib}
\end{document}